\documentstyle[11pt,epsf,graphicx,subfigure]{article}
\begin{document}
\baselineskip 100pt
\renewcommand{\baselinestretch}{01.35}
\renewcommand{\arraystretch}{0.666666666}
\large
\parskip.2in
\newcommand{\hs}{\hspace{1mm}}
\newcommand{\nhat}{\mbox{\boldmath$\hat n$}}
\newcommand{\cmod}[1]{ \vert #1 \vert ^2 }
\newcommand{\mod}[1]{ \vert #1 \vert }
\newcommand{\pr}{\partial}
\newcommand{\r}{\rho}
\newcommand{\fr}{\frac}
\newcommand{\ie}{{\em ie }}
\newcommand{\th}{\theta}
\newcommand{\p}{\varphi}
\newcommand{\x}{\xi}
\newcommand{\xb}{\bar{\xi}}
\newcommand{\Vd}{V^{\dagger}}
\newcommand{\Ref}[1]{(\ref{#1})}

\title{\hbox{{\Large{\bf Spherically symmetric solutions of the 
6th order SU(N) Skyrme models.}}}}

\author{
I. Floratos\thanks{e-mail address: Ioannis.Floratos@durham.ac.uk}
\, and B.M.A.G. Piette\thanks{e-mail address:
B.M.A.G.Piette@durham.ac.uk}
\\ Department of Mathematical Sciences,University of Durham, \\
 Durham DH1 3LE, UK\\
}\date{June 2001}

\maketitle

\begin{abstract}
Following the construction described in \cite{sun}, we use 
the rational map ansatz to construct analytically some topologically 
non-trivial solutions of the generalised $SU(3)$ Skyrme model defined by adding
a sixth order term to the usual Lagrangian. These solutions are radially
symmetric and some of them can be interpreted as bound states of Skyrmions.
The same ansatz is used to construct low-energy configuration of the $SU(N)$ 
Skyrme model.
\end{abstract}

%%%%%%%%%%%%%%%%%%%%%%%%%%%%%%%%%%%%%%%%%%%%%%%%%%%%%%%%%%%%%%%%%%%%%%%%%%%%%%
\section{Introduction}

The Skyrme model \cite{Skyr1} is widely accepted as an effective theory to 
describe the low-energy properties of
nucleons. It was indeed shown \cite{Hooft,Witt1,Witt2} that in the large 
$N_c$ limit, the Skyrme model is the low-energy limit of QCD. 
The classical static solutions of the model describe
the bound states of nucleons and every configuration is characterised 
by a topological charge which following Skyrme's idea, is interpreted as
the baryon charge. 

The Skyrme model can be used to predict the properties of the nucleons within 
$10$ to $20\%$ \cite{Witt1,Witt2}. To improve these phenomenological 
predictions various extensions of the models have been proposed mostly by 
adding higher order terms \cite{Jackson,Marl1,Marl4,Marl5}
or extra fields \cite{HA} to the Lagrangian.
 
The study of the classical solutions of the Skyrme model has been done mostly
using numerical methods, but recently Houghton et al. \cite{Manton}
showed that the classical solutions of the $SU(2)$ model can be well 
approximated by using an ansatz that involves the harmonic maps from $S^2$ to 
$S^2$. The harmonic map describes the angular distribution of the solution 
while a profile function describes its radial distribution. 
This construction was later generalised \cite{spherical} for the $SU(N)$ 
model using harmonic maps from $S^2$ to $CP^{N-1}$. Moreover, it was shown that
using a further generalisation of this ansatz one can construct exact  
spherically symmetric solutions of the $SU(N)$ Skyrme model.

The same method was also used in \cite{Hans} to construct solutions of 
another $SU(N)$ 4th order Skyrme model.
In this paper, we use the same generalised ansatz to construct solutions
of the sixth order $SU(3)$ Skyrme model and low-energy configurations of the 
$SU(N)$ models defined in \cite{FP1}.

%%%%%%%%%%%%%%%%%%%%%%%%%%%%%%%%%%%%%%%%%%%%%%%%%%%%%%%%%%%%%%%%%%%%%%%%%%%%%%
\section{The sixth order Skyrme model}

The Skyrme model is described by an $SU(N)$ valued field $U(\vec{ x},t)$
which, to ensure finiteness of the energy,  is required to satisfy the 
boundary condition $U \rightarrow I$ as $|\vec{x}| \rightarrow \infty$, 
where $I$ is the unit matrix. This boundary condition implies that the three 
dimensional Euclidean space on which the model is defined 
can be compactified into $S^3$ and as a result, the Skyrme field $U$ 
corresponds to mappings from $S^3$ into $SU(N)$. 
As $\pi_3(SU(N)) = Z$ each configuration is characterised by its winding 
number, or topological charge, which can be obtained explicitly by evaluating 
the expression
\begin{eqnarray}
\label{topcharge}
B=\frac{1}{24\pi^2}\int_{R^3}d\vec{x}\hspace{1mm}^3\hs
\varepsilon^{ijk}\hspace{1mm}Tr(R_i \hs R_j \hs R_k ),
\end{eqnarray}
where $R_{\mu}=(\partial_{\mu}U)U^{-1}$ is the right chiral current.
Skyrme's ideas was to interpret the winding number associated with these 
topologically non-trivial mappings as the baryon charge.

The generalised sixth order Skyrme model is defined by the Lagrangian
\begin{eqnarray}
\label{energy}
E=-\frac{1}{12\pi^2}\hspace{1mm}\int d\vec{x }\hspace{1mm}^3
  \left(\frac{1}{2} \hspace{1mm}Tr R_{i}^2+
        \frac{1-\lambda}{16}\hspace{1mm} Tr[R_{i},R_{j}]^2+
      \frac{1}{96}\lambda\hspace{1mm} Tr[R_{i},R_{j}][R_{j},R_{k}][R_{k},R_{i}]
  \right),
\end{eqnarray}
where this parametrisation of the model is chosen such that 
$\lambda \in [0,1]$ 
is a mixing parameter between the Skyrme term and the sixth order term: 
when $\lambda=0$ the model reduces to the usual pure Skyrme model while for 
$\lambda = 1 $  the Skyrme term vanishes and the model reduces to what we 
refer to in what follows as the pure Sk6 model. 

The Euler-Lagrange equations derived from \Ref{energy} for the static
solutions are given by
\begin{eqnarray}
\label{eqnR}
\partial_i\left(R_i-\frac{1}{4}(1-\lambda)\Big[R_j,[R_j,R_i]\Big]-
\frac{1}{16}\lambda\Big[R_j,[R_j,R_k][R_k,R_i]\Big]\right)=0.
\end{eqnarray}
and the following inequality holds for every configuration
\begin{eqnarray}
\label{topchargein}
\tilde{E} \ge \sqrt{1-\lambda} B.
\end{eqnarray}

The multi-Skyrmion solutions of the $SU(2)$ Skyrme model have been studied 
in \cite{FP1} where it was shown that they have the same symmetry as the 
pure Skyrme model. It was also shown that the harmonic map ansatz gives 
a good approximation to the solutions.

In the next section we describe the harmonic map ansatz. In the third section
we prove that due to a constraint coming from the sixth order term, 
the multi-projector harmonic map ansatz provides exact solutions
only for the $SU(3)$ generalised model. We then show that one can 
nevertheless use the ansatz to construct low-energy configurations of the 
$SU(N)$ models. In the fourth section we look at these configurations 
for the $SU(4)$ model, while in the last section we look at some special 
ansatz for the $SU(N)$ model.

%%%%%%%%%%%%%%%%%%%%%%%%%%%%%%%%%%%%%%%%%%%%%%%%%%%%%%%%%%%%%%%%%%%%%%%%%%%%%%
\section{Harmonic map ansatz}

The rational map ansatz, introduced by Houghton et al. \cite{Manton} is a 
generalisation of the hedgehog ansatz found by Skyrme \cite{Skyr1}, to 
approximate multi-Skyrmion solution of the $SU(2)$ model. The ansatz was later 
generalised by Ioannidou et al. \cite{sun}
to approximate solutions of the $SU(N)$ Skyrme model using harmonic maps
from $S^2$ into $CP^{N-1}$. This generalised ansatz is given by   
\begin{eqnarray}
\label{genansatz}
U(r,\theta, \varphi)&=&e^{2if(r)(P(\theta, \varphi)-{\it I}/N)}\nonumber\\
        &=&e^{-2if(r)/N}\left({\it I}+(e^{2if(r)}-1)P(\theta, \varphi)\right)
\end{eqnarray}
where $r, \theta$ and $\varphi$ are the usual polar coordinates.
The profile function $f(r)$ must satisfy the boundary conditions 
$f(0) = \pi$ and  $\lim_{r\rightarrow \infty} f(r) = 0$ and  
$P(\theta, \varphi)$ is a projector in $C^N$ which must be a harmonic 
map from $S^2$ into $CP^{N-1}$ or equivalently a classical solution of the
2 dimensional $CP^{N-1}$ $\sigma$ model. These solutions are well known 
\cite{Din,Wojtek} and to construct them it is convenient to introduce the 
complex coordinate
$\xi=\tan(\theta/2)e^{i\varphi}$ which corresponds to the stereographic
projection of the unit sphere onto the complex plane.

In these coordinates, $P$ must satisfy the equation
\begin{equation}
\label{selfdual}
 P \frac{\partial P}{\partial \xi} = 0.
\end{equation}
and the solutions of that equation are given by any projector of the form
\begin{equation}
\label{projector}
P(f)=\frac{h \otimes h^\dagger}{|h|^2}
\end{equation}
where $h \in C^N$ is holomorphic
\begin{equation}
 \frac{\partial h}{\partial \bar{\xi}} = 0.
\end{equation}

The topological charge for the ansatz \Ref{genansatz}, with the prescribed 
boundary conditions for $f(r)$,  is given by the winding number of the
$S^2 \rightarrow CP^{N-1}$. This winding number which is itself given by 
the degree of the harmonic function $h$ \cite{Din,Wojtek} which must then be 
a rational function of $\xi$.

To approximate a solution, one plugs the ansatz \Ref{genansatz} into the 
energy \Ref{energy} and notices that if $P$ satisfies \Ref{selfdual},
the integration over the polar angles and the radius decouple. 
One then has to minimise the integral over the polar angles  of an expression
which depends only on $P$. Taking for $P$ the most general harmonic map of 
the desired degree, one then has to find the parameters of the general map
which minimise that integral. Having done this, the profile function $f$ is
obtained by solving the Euler Lagrange equation derived from the effective 
energy.

A special case of this construction is the so-called hedgehog ansatz for
the $SU(2)$ model corresponding to one Skyrmion. In this case, 
we have $h = (1,\xi)^t$ and after inserting \Ref{projector} into \Ref{energy} 
the energy reduces to 
\begin{eqnarray}
\label{energyans}
E={1 \over 3 \pi} \int dr \hspace{3mm} 
 ( f_r^2 \,r^2 + 2\, \sin^2 f\, (1+(1-\lambda) f_r^2) 
      +(1-\lambda) \frac{\sin^4 f}{r^2} +\lambda\frac{\sin^4 f}{r^2}\,f_r^2)
\end{eqnarray}
and the equation for $f$ is given by
\begin{eqnarray}
\label{eqng}
f_{rr} \left( 1+2\,(1-\lambda) \,\frac{\sin^2 f}{r^2}
          + \,\lambda\,\frac{\sin^4 f }{r^4}
      \right)
  +\frac{2}{r}\,f_r\,\left( 1-\,\lambda\,
                \frac {\sin^4 f }{r^4}\right ) &&\nonumber
\\ \hspace{30mm}+ \,\frac{\sin2g }{r^2} 
   \left ( (1-\lambda) f_{r}^{2}-1
           +\frac{\sin^2 f}{r^2}(\lambda\,f_{r}^{2}-1+\lambda)
   \right)=0.
\end{eqnarray}
This actually corresponds to an exact solution of the model and it is radially 
symmetric. In Figure 1 we present the $\lambda$ dependence of the energy and 
in Figure 2 we show the profile function $f$ and the energy density for the 
pure Skyrme model, $\lambda = 0$, and the pure Sk6 model, $\lambda = 1$.

\vskip 5mm
\begin{figure}[htbp]
\unitlength1cm \hfil
  \includegraphics[width=8cm]{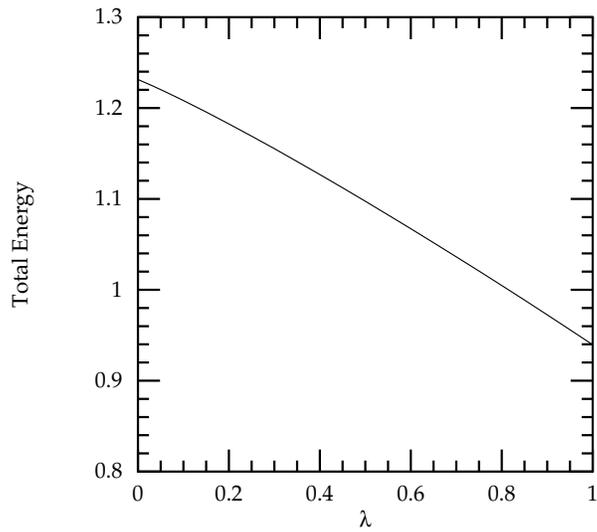}
\caption{Total energy of the 1 Skyrmion solution.}
\end{figure}

\vskip 5mm
\begin{figure}[htbp]
\unitlength1cm \hfil
  \includegraphics[width=7cm]{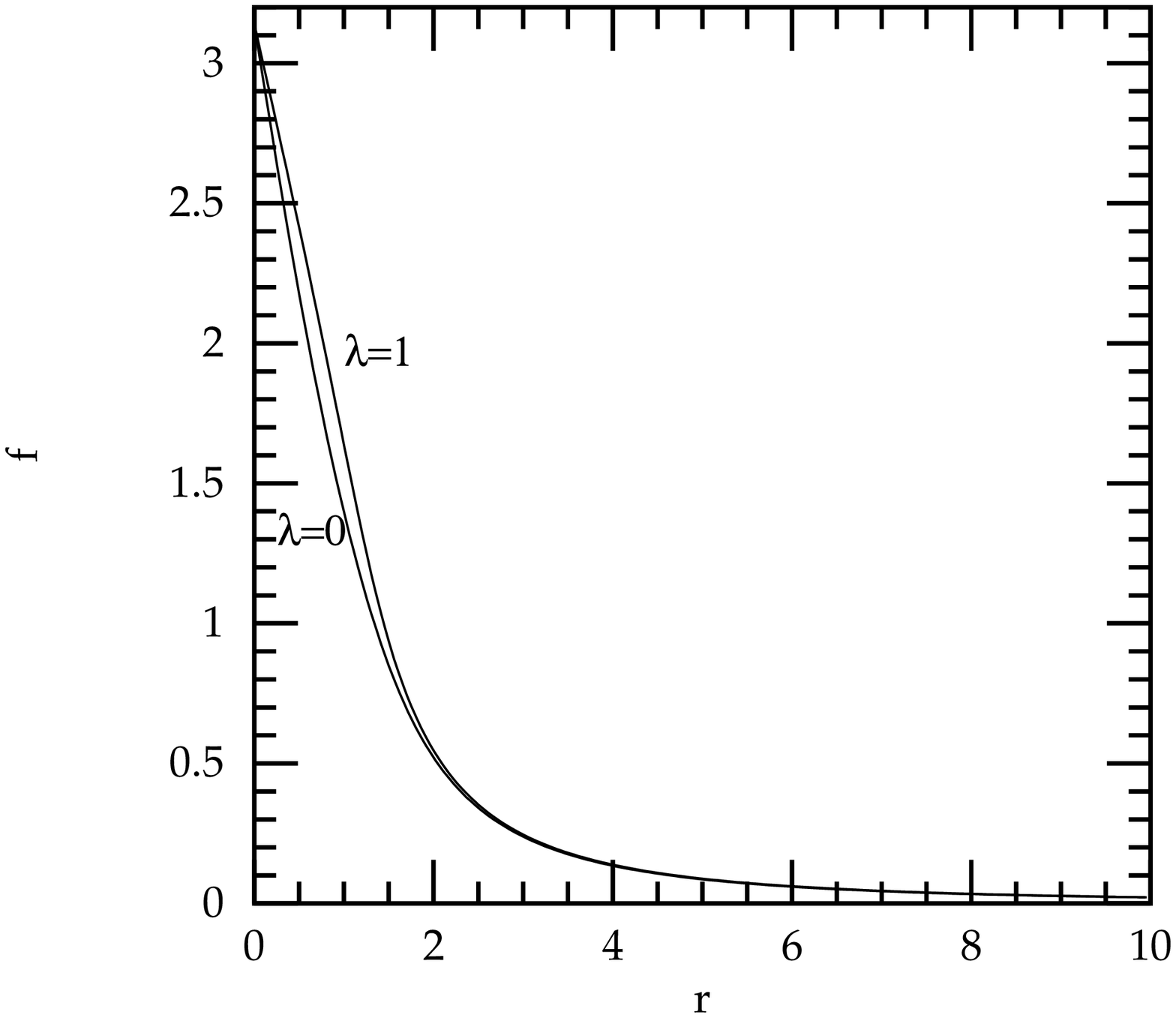}
  \includegraphics[width=7cm]{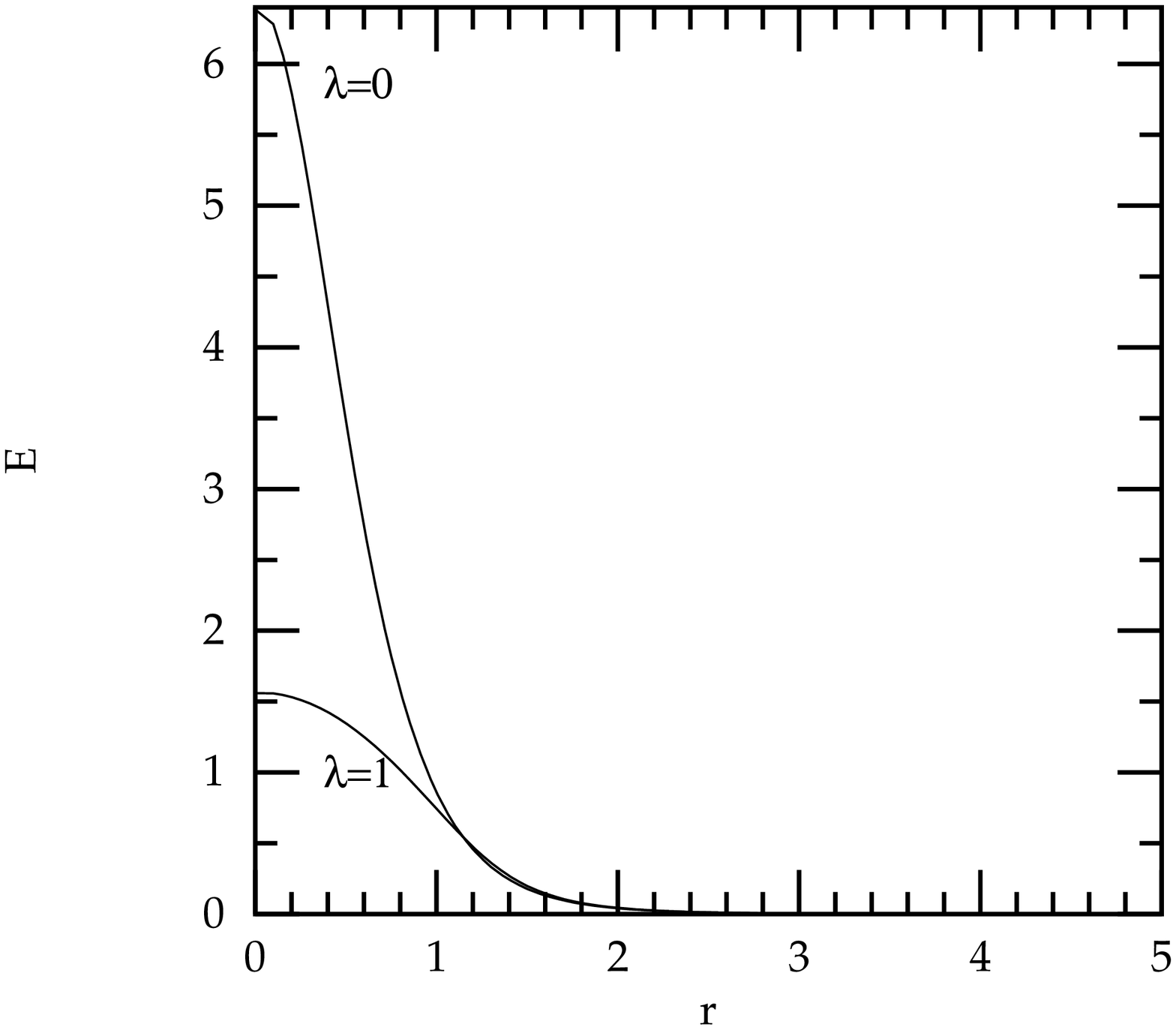}
\caption{Function profile $f$ and energy density for the 1 Skyrmion solution 
of the pure Skyrme model, $\lambda = 0$, and the pure Sk6 model, 
$\lambda = 1$. .}
\end{figure}

%%%%%%%%%%%%%%%%%%%%%%%%%%%%%%%%%%%%%%%%%%%%%%%%%%%%%%%%%%%%%%%%%%%%%%%%%%%%%%
\section{Spherically symmetric solutions for the $SU(N)$ model}

In this section we will follow the construction described in \cite{sun},
to attempt to construct solutions of the extended $SU(N)$ Skyrme model using a 
generalisation of the harmonic map ansatz \Ref{genansatz}. 

To build the new ansatz we need to introduce an operator $P_+$ which acts
on any complex vector $u \in C^N$ and is defined as
\begin{eqnarray}
P_+ u=\pr_\xi u- u\,\fr{u^\dagger \,\pr_\xi u}{|u|^2}.
\end{eqnarray}
Taking a holomorphic vector $h(\xi)$ we then define $P_0^+h\,=\,h$ and by
induction $V_k = P^k_+ h = P_{+}(P^{k-1}_+ h)$. These $N$ vectors are 
mutually orthogonal \cite{Wojtek} and so the corresponding projectors 
\begin{eqnarray}
P_k\,=\,P(P^k_+ h)\, = \, {P^k_+ h (P^k_+ h)^\dagger \over \cmod{P^k_+ h}} 
\hs \hs k=0,\dots,N-1,
\end{eqnarray}
satisfy the orthogonality relations 
\begin{eqnarray}
\label{projprop}
P_k P_j &=& \delta_{ij} P_k \nonumber \\
\sum_{k=0}^{N-1}P_k &=& 1
\end{eqnarray}
as well as other properties discussed in detail in \cite{sun}.

The generalised harmonic map ansatz is then defined as
\begin{eqnarray}
\label{ansatz}
U&=&exp\{ig_0
(P_0-\fr{I}{N})+ig_1(P_1-\fr{I}{N})-\dots+ig_{N-2}(P_{N-2}-\fr{I}{N})\}
\nonumber\\
&=&e^{-ig_0/N}(I+A_0P_0)\,\,e^{-ig_1/N}(I+A_1P_1)\,\dots\,
\,e^{-ig_{N-2}/N}(I+A_{N-2}P_{N-2})
\end{eqnarray}
where $g_k(r)$ are $N-1$ profile functions and $A_k=e^{ig_k}-1$. Moreover for 
the ansatz to be well defined, the profile functions $g_k(r)$ must be a
a multiple of $2 \pi$ at the origin and at infinity. 

To proceed with our construction, it is convenient to rewrite the 
Euler-Lagrange equations of the model \Ref{eqnR} using the usual 
spherical coordinates 
\begin{eqnarray}
\label{eqnpolar}
&& \pr_r\,
\Bigg\{ r^2 R_r+\fr{1-\lambda}{4}\left(A_{\th r \th}
       +\fr{1}{\sin^2\th}A_{\p r \p}\right)+\fr{1}{16}\lambda \,
 \left[\fr{1}{\sin^2 \th}\left(B_{\th \th \p r \p} +B_{\p \p \th r \th}\right)
 \right]
\Bigg\}+ 
\nonumber \\ 
&& \fr{1}{\sin\th}\,\pr_\th 
\Bigg\{\sin\th 
\left[R_\th +\fr{1-\lambda}{4}\left(A_{r\th r}
      +\fr{1}{r^2\sin^2\th}A_{\p \th \p}\right)
\right]
+\fr{\lambda}{16\, r^2 \sin^2\th}
 \left( B_{r r \p \th \p}+B_{\p \p r \th r}\right)
\Bigg\}
\nonumber \\ 
&& +\fr{1}{\sin^2\th}\,\pr_\p \Bigg\{ R_\p +\fr{1-\lambda}{4}\left(A_{r\p r}
 +\fr{1}{r^2}A_{\th \p \th}\right)
 +\fr{\lambda}{16\, r^2} \left( B_{r r \th \p \th}
                           +B_{\th \th r \p r}\right)
                         \Bigg\}=0
\end{eqnarray}
where $A_{jij}\equiv \Big[R_j,\,[R_i,\,R_j]\Big]$ and
$B_{jjkik}\equiv \Big[R_j,\,[R_j,R_k]\,[R_i,R_k]\Big]$ . It is
fairly easy to show that
\begin{eqnarray}
R_r=i\sum_{j=0}^{N-2}\dot g_j\left(P_j-\fr{I}{N}\right),
\end{eqnarray}
where $\dot g_j$ is the derivative of $g_j(r)$ with respect to
$r$. Using the complex coordinates $\xi$ and $\bar{\xi}$
introduced before we have
\begin{eqnarray}
R_\xi=\sum_{i=1}^{N-1}\left[e^{i(g_i-g_{i-1})}-1\right]\fr{V_i\,V^{\dagger}_{i-1}}
{|V_{i-1}|^2}
\end{eqnarray}
and the derivatives with respect to $\th$ and $\p$ are given by
\begin{eqnarray}
  \pr_{\th}=\fr{1+|\xi|^2}{2\sqrt{|\xi|^2}}
         \left(\xi\,\pr_\xi\,+\bar\xi\,\pr_{\bar\xi}\right),
\hspace{10mm}
  \pr_{\varphi}=i\left(\xi \,\pr_\xi\,-\bar\xi\,\pr_{\bar\xi}\right).
\end{eqnarray}
Substituting the  above into  equations \Ref{eqnpolar} we get
\begin{eqnarray}
\label{eqnxi}
&&\pr_r\left[r\sp2R_r+ (1-\lambda)\fr{(1+|\x|^2)^2}{8}
                      \left(A_{\xb\,r\,\x}+A_{\x\,r\,\xb}\right)
      \right]
   +\fr{(1+|\x|^2)^2}{2}\left((R_{\xb})_\x+(R_{\x})_{\xb}\right)+\nonumber\\
&& (1-\lambda)\fr{(1+|\x|^2)^3}{8r^2}
       \left(\x\, A_{\x\,\x\,\xb}-\xb\,A_{\xb\,\x\,\xb} \right)
   +(1-\lambda)\fr{(1+|\x|^2)^4}{16r^2} 
     \left(\left[A_{\x\,\x\,\xb}\right]_{\xb}
          -\left[A_{\xb\,\x\,\xb}\right]_{\x}
    \right)\nonumber\\
&& +(1-\lambda)\fr{(1+|\x|^2)^2}{8} 
     \left(\left[A_{r\,\xb\,r}\right]_\x
           +\left[A_{r\,\x\,r}\right]_{\xb}
    \right)
    +\fr{\lambda}{16} \Bigg\{\pr_r
          \left[\fr{(1+|\x|^2)^4}{4}
               \left(B_{\xb\,\x\,\xb\,r\,\xb}-B_{\x\,\x\,\xb\,r\,\xb}
               \right)
          \right]+\nonumber\\ 
&& \fr{(1+|\x|^2)^2}{4r^2}
     \left(\pr_{\xb}
       \left[(1+|\x|^2)^2\, B_{r\,r\,\x\,\x\,\xb}\right]- \pr_\x
       \left[(1+|\x|^2)^2\,B_{r\,r\,\xb\,\x\,\xb}\right]
    \right)+\nonumber\\
&& \fr{(1+|\x|^2)^2}{2|\x|^2\,r^2}
     \left(\x\pr_{\x}\left[\fr{(1+|\x|^2)^2}{4|\x|^2} 
                          (-\x\x\x B_{\x\,\x\,r\,\x\,r})
                    \right]
            +\xb\pr_{\xb}\left[\fr{(1+|\x|^2)^2}{4|\x|^2}
                               (-\xb\xb\xb B_{\xb\,\xb\,r\,\xb\,r})
                        \right]
    \right)+\nonumber\\
&& \fr{(1+|\x|^2)^2}{8r^2}\Bigg(
         \pr_\x\Big[(1+|\x|^2)^2(B_{\x\,\xb\,r\,\xb\,r}
                    +B_{\xb\,\x\,r\,\xb\,r}-B_{\xb\,\xb\,r\,\x\,r})
               \Big] +\nonumber \\
& & \hspace{35mm}
      \pr_{\xb}\Big[(1+|\x|^2)^2(-B_{\x\,\x\,r\,\xb\,r}+B_{\x\,\xb\,r\,\x\,r}+
                    B_{\xb\,\x\,r\,\x\,r})
               \Big]
    \Bigg)\Bigg\}=0.
\end{eqnarray}
In \cite{sun} it is shown that if one takes the special holomorphic 
vector
\begin{equation}
\label{V}
  V_0\, =\, h\, =\, (h_0, h_1, \dots, h_{N-1})^t
\end{equation}
where
\begin{equation}
\label{fk}
  h_k = \xi^k \sqrt{C_{k}^{N-1}}
\end{equation}
and where $C_{k}^{N-1}$ denotes the binomial coefficients, then the
terms in \Ref{eqnxi} coming from the usual Skyrme model, {\it i.e.} all 
the terms except the ones proportional to $\lambda/16$, are all proportional 
to $P_i-P_{i-1}$ and $P_i-\fr{I}{N}$. Using \Ref{projprop} one can get rid of 
the projector $P_{N-1}$ and \Ref{eqnxi} will then be the sum of the $N-1$ 
terms $P_i-\fr{I}{N}$ for $i = 0 \dots N-2$, with coefficients that depend 
only on $r$. This implies that the equations for the Skyrme model reduce to 
$N-1$ ordinary 
differential equations for the profile functions $g_i$ and their solutions,
if they exist, will provide us with  exact solutions of the $SU(N)$ Skyrme 
model.

In what follows we will show that the angular dependence of the terms 
proportional to $\lambda$ in \Ref{eqnxi}, {\it i.e.} the terms coming from 
the sixth order term, is also coming exclusively from the projectors 
$P_{i}-\frac{I}{N}$ or $P_{i}-P_{i-1}$ but that we have to impose an extra 
constraint on the profile functions $g_i$. 

We start by noting that 
\begin{eqnarray}
& &
[R_\x,R_{\xb}] = -\sum_{i=1}^{N-1}a^{2}_i\,\fr{|V_i|^2}{|V_{i-1}|^2}
                 \left(\fr{V_{i}\Vd_{i}}{|V_{i}|^2}
                      -\fr{V_{i-1}\Vd_{i-1}}{|V_{i-1}|^2}
                \right)\\
& & [R_r,R_{\x}] = i \sum_{i=1}^{N-1} 
                 \left(\dot g_i a_i- \dot g_{i-1} a_i\right)\,
          \fr{V_{i}\Vd_{i-1}}{|V_{i-1}|^2} = 
                 \sum_{i=1}^{N-1} K_i \fr{V_{i}\Vd_{i-1}}{|V_{i-1}|^2} \\ 
& & [R_r,R_{\xb}] = i \sum_{i=1}^{N-1} 
                 \left(\dot g_i a_i- \dot g_{i-1} a_i\right)\,
                       \fr{V_{i-1}\Vd_{i}}{|V_{i-1}|^2}
                  = \sum_{i=1}^{N-1} K_i \fr{V_{i-1}\Vd_{i}}{|V_{i-1}|^2}
\end{eqnarray}
where $a_i=e^{i(g_i-g_{i-1})}-1$. It is then straightforward to
check that 
\begin{eqnarray}
B_{\xb\x\xb r \x}-B_{\x\x\xb r \xb} = \sum_{i=1}^{N-1}
      \left(b_i\fr{|V_{i-1}|^2}{|V_{i-2}|^2}\fr{|V_{i}|^2}{|V_{i-1}|^2}
           +c_i\fr{|V_{i}|^4}{|V_{i-1}|^4}
           +d_i\fr{|V_{i+1}|^2}{|V_{i}|^2}\fr{|V_{i}|^2}{|V_{i-1}|^2}
     \right)
     \left(P_i-P_{i-1}\right)
\end{eqnarray}
where $b_i, c_i$ and $d_i$ are functions of $g_k$ only.
However, as shown in \cite{sun}, if $V_0$ is given by 
\Ref{V} and \Ref{fk} then
$\fr{|V_{i}|^2}{|V_{i-1}|^2}\propto(1+|\x|^2)^{-2}$ and thus
\begin{eqnarray}
\fr{(1+|\x|^2)^4}{4}\left(B_{\xb\,\x\,\xb \,r \,\x}-B_{\x\,\x\,\xb
\,r \,\xb}\right)\propto\left(P_i-P_{i-1}\right).
\end{eqnarray}
Furthermore,  we have 
\begin{eqnarray}
B_{r\,r\,\x\,\x\,\xb}=i\sum_{i=1}^{N-1}
               \left(e_i\fr{|V_{i}|^2}{|V_{i-1}|^2}+
                     s_i\fr{|V_{i-1}|^2}{|V_{i-2}|^2}
              \right) \fr{V_{i}\Vd_{i-1}}{|V_{i-1}|^2}
\end{eqnarray}
with $e_i= e(g_i)$ and $s_i=s(g_i)$. But in
equation \Ref{eqnxi} this term  appears as
\begin{eqnarray}
\label{eqn48}
\pr_{\xb}\left[(1+|\x|^2)^2 B_{r\,r\,\x\,\x\,\xb}\right] =
      2\x \,(1+|\x|^2) B_{r\,r\,\x\,\x\,\xb}+
            (1+|\x|^2)^2\,\,\pr_{\xb}\, (B_{r\,r\,\x\,\x\,\xb}).
\end{eqnarray}
Since $\pr_{\xb}\,\fr{|V_{i}|^2}{|V_{i-1}|^2} \propto -2\x\,(1+|\x|^2)^{-3}$ 
the only  parts of \Ref{eqn48} that are non zero are the ones that involve 
the derivatives of $\fr{V_{i}\Vd_{i-1}}{|V_{i-1}|^2}$ with respect to $\xb$.
Since it can be shown that the latter are proportional to 
$\sum_{i-1}^{N-1}C_i (1+|\x|^2)^{-2} \left(P_i-P_{i-1}\right)$ where 
$C_i=C(g_i)$, then one sees that the term that involves 
$B_{r\,r\,\x\,\x\,\xb}$ in \Ref{eqnxi} is proportional to 
$\left(P_i-P_{i-1}\right)$.

Using similar arguments, it is easy to check that the terms
involving $\,\, B_{r\,r\,\xb\,\x\,\xb}$, $\,
B_{\x\,\xb\,r\,\xb\,r}$, $\, B_{\xb\,\x\,r\,\xb\,r}$,
$\,B_{\xb\,\xb\,r\,\x\,r}$, $\, B_{\x\,\x\,r\,\xb\,r}$, $\,
B_{\x\,\xb\,r\,\x\,r}$ and $\, B_{\xb\,\x\,r\,\x\,r}\,$ 
factorise in the same way.

There are a few terms in \Ref{eqnxi} which we still have to consider. 
They involve the expressions
\begin{eqnarray}
& & 
\label{B4_1}
B_{\x\,\x\,r\,\x\,r}=\sum_{i=3}^{N-1} \left(a_i
                     K_{i-1}K_{i-2}-a_{i-2} K_{i}K_{i-1}\right)
                     \fr{V_{i}\Vd_{i-3}}{|V_{i-3}|^2} \\ & &
\label{B4_2}
    B_{\xb\,\xb\,r\,\xb\,r}=\sum_{i=3}^{N-1} \left(a_i
                     K_{i-1}K_{i-2}-a_{i-2} K_{i}K_{i-1}\right).
                     \fr{V_{i-3}\Vd_{i}}{|V_{i-3}|^2}
\end{eqnarray}
where $K_i=i\left(\dot g_i a_i- \dot g_{i-1} a_i\right)$.
It is clear that these terms will always give a $\x,\xb$
dependence besides the projectors $P_i$ and hence, if we want
\Ref{eqnxi} to reduce to $N-1$ equations that involve only
the profile functions $g_i$ then we have to make sure that \Ref{B4_1} and
\Ref{B4_2} vanish \ie we must impose the conditions 
\begin{eqnarray}
& & a_i K_{i-1}K_{i-2}-a_{i-2} K_{i}K_{i-1}=0
\hspace{5mm}\Leftrightarrow \hspace{5mm} \dot g_i=\dot g_{i-2}.
\end{eqnarray}
This last constraint which is a result of the addition of the sixth order 
term, implies that we can only consider two profile functions $g_0$ and $g_1$ 
and that we should thus have only two equations. Unfortunately we have
$N-1$ equations which are not compatible with each other. From this we see 
that the ansatz \Ref{genansatz} will provide exact solutions of the 
generalised Skyrme model for the $SU(2)$ and the $SU(3)$ model only. 
For larger values of $N$, the ansatz will nevertheless give some low-energy 
radially symmetric configurations. The $SU(2)$ case is nothing but the usual 
hedgehog ansatz and we will focus on the solutions of the $SU(3)$ model in 
the next section.

In order to derive the equations for the profile functions, it is convenient
to write the energy density of the model in terms of $(\x,\xb)$:
\begin{eqnarray}
\label{eqnans}
& &E=-\fr{i}{12\pi^2}\,\,\int r^2 dr\,d\x d\xb
      \,\,\,\,Tr\,\,\Bigg(\fr{1}{(1+|\x|^2)^2}R^{2}_r+ \fr{1}{r^2} |R_\x|^2 
      +\fr{1-\lambda}{4r^2}[R_r,R_\x][R_r, R_{\xb}] \nonumber \\ 
& & \hspace{10mm} 
  -(1-\lambda)\fr{(1+|\x|^2)^2}{16r^4}[R_{\xb},R_\x]^2+\lambda\,
        \fr{(1+|\x|^2)^2}{64r^4}\Bigg[[R_r,R_{\xb}]\, ,\,
           [R_r,R_{\x}]\Bigg][R_\x,R_{\xb}]\Bigg).
\end{eqnarray}
Defining 
\begin{eqnarray}
\label{defFi}
F_i &=& g_i - g_{i+1}\qquad \mbox{for}\quad i=0,\dots,N-3,\nonumber\\
F_{N-2} &=& g_{N-2}
\end{eqnarray}
as well as 
$W_i=\fr{|V_i|^2}{|V_{i-1}|^2}\left(1-\cos(F)\right)$ 
and
$W_{N-1}=\fr{|V_{N-1}|^2}{|V_{N-2}|^2}\left(1-\cos(g)\right)$ 
the terms in the above expression can be rewritten as
\begin{eqnarray}
&&Tr\,R_{r}^2=\fr{1}{N}\left(\sum_{i=0}^{N-2}\dot g_i\right)^2\,
                       -\,\sum_{i=0}^{N-2}\dot g_{i}^2 \,\, , \\
&&Tr\,|R_\x|^2\,=\,-2\sum_{i=1}^{N-1} W_i\,\, ,\\
&&Tr\,[R_r,R_\x][R_r,R_{\xb}]=-2\sum_{k=1}^{N-1} W_k\,\,\dot F_{k-1}^2, \\ 
&&Tr\,[R_{\xb},R_\x]^2=4\left(W_{1}^2 +
               \sum_{i=1}^{N-2} (W_i-W_{i+1})^2+W_{N-1}^2\right),
\end{eqnarray}
\begin{eqnarray}
&&  Tr \,\Bigg[[R_r,R_{\xb}]\, ,\, 
    [R_r,R_{\x}][R_\x,R_{\xb}]\Bigg] = 4 \Big(\dot F_{0}^2 W_{1}^2
               +\sum_{i=1}^{N-2}\left(\dot F_{i-1}W_i- \dot F_{i}W_{i+1}
                               \right)^2
               + \dot F_{N-2}^2 W_{N-1}^2 \Big). \nonumber\\
\end{eqnarray}
In \cite{sun} it was shown that 
\begin{equation}
\label{normV}
\fr{|V_k|^2}{|V_{k-1}|^2}=k(N-k)(1+|\x|^2)^{-2}
\end{equation} 
and from this we see that
all the terms in \Ref{eqnans} are proportional to $(1+|\x|^2)^{-2}$ and that 
after integrating out the angular dependence the energy reduces to

\begin{eqnarray}
\label{SunEnergy}
E\!\!\!\!&=&\!\!\!\frac{1}{6\pi}\int\!\! r^{2} dr\Bigg\{-{1\over N}
\left(\sum_{i=0}^{N-2}\dot g_i\right)^2+\sum_{i=0}^{N-2}\dot g_i^2
  +{2\over r^2}\sum_{k=1}^{N-1}Z_k
  +{(1-\lambda)\over2r^2}\sum_{k=1}^{N-1}
   \left(\dot g_k-\dot g_{k-1}\right)^2 Z_k
\nonumber\\
&&\hspace{17mm}+{(1-\lambda)\over 4r^4}
  \left(Z_1^{2}+\sum_{k=1}^{N-2}(Z_k-Z_{k+1})^2+Z_{N-1}^2\right) 
\nonumber \\
&&\hspace{17mm}+{\lambda\over 16r^4}
 \left(\dot F_{0}^{2}\,Z_1^{2}+\sum_{k=1}^{N-2}(\dot F_{k-1}\,Z_k-
  \dot F_{k}\,Z_{k+1})^2+\dot F_{N-2}^2\,Z_{N-1}^2\right)\Bigg\},
\end{eqnarray}
where $Z_k=k(N-k)(1-\cos(F_{k-1}))$.

In \cite{sun} the fields $F_i$ defined by \Ref{defFi} were used, and very 
special solutions were obtained by taking $F_0 = F_1 =\dots =F_{N-2}$. 
It was observed that when $F_i(0) = 2\pi$ and $F_i(\infty) = 0$ this solution 
of the $SU(N)$ 
pure Skyrme model has a topological charge $B=\frac{N}{6}(N^2-1)$ and has an 
energy equal exactly to $\frac{N}{6}(N^2-1)$ times the energy of the single 
Skyrmion solutions. It is easy to show that, if one uses the same ansatz 
for the sixth order Skyrme model, the profile $f = F_0/2$ satisfies the 
hedgehog profile equation \Ref{eqng} and the energy of the configuration
is given by  $E(\lambda) = 4 E_0(\lambda)$ where  $E_0(\lambda)$ is the energy 
of the hedgehog solution for the generalised model. These configurations are 
not exact solutions, except for the $SU(3)$ model.

To consider the most general ansatz, one can derive from \Ref{SunEnergy}  
the following equations for the profile functions $F_l$, $l=0,..(N-2)$.
 
\begin{eqnarray}
&&-{2(l+1)\over N}\sum_{i=0}\sp{N-2}(i+1)\ddot F_i+2\sum_{k=0}\sp{l}
\sum_{i=k}\sp{N-2}\ddot F_i+\frac{(1-\lambda)}{r^2}\ddot
F_l(l+1)(N\!-\!l\!-\!1)(1-\cos F_l)+
\nonumber\\
\label{general_N}
&&\!{2\over r}\left(\!-{2(l\!+\!1)\over N}\sum_{i=0}\sp{N-2}(i+1)\dot F_i\!
  +\!2\sum_{k=0}\sp{l}\left(\sum_{i=k}\sp{N-2}\dot F_i\right)\right)
  +{(1-\lambda)\over 2r\sp2}\dot F_l\sp2(l\!+\!1)(N\!-\!l\!-\!1)\sin F_l\!+
\nonumber\\
&&-{2\over r\sp2}\,(l+1)(N-l-1)\,\sin F_l- 
 {(1-\lambda)\over r\sp4}\,(l+1)\sp2(N-l-1)\sp2\left(1-\cos F_l\right)\sin F_l+
\nonumber\\
&&{(1-\lambda)\over 2r\sp4}(l\!+\!1)(N\!-\!l\!-\!1)\sin F_l 
     \left[l(N\!-\!l)(1-\cos F_{l-1})
     +(l\!+\!2)(N\!-\!l\!-\!2)(1-\cos F_{l+1})\right]
\nonumber \\
&&+\frac{\lambda}{8r^4}\Bigg\{2\,\ddot F_l (l+1)^2 (N-l-1)^2(1-\cos F_l)^{2}
        -\ddot F_{l-1}l(l+1) (N-l)(N-l-1)
\nonumber \\
&&(1-\cos F_{l-1})(1-\cos F_{l})-\ddot F_{l+1}(l+1)(l+2)(N-l-1)(N-l-2)
\nonumber\\ 
&&(1-\cos F_{l})(1-\cos F_{l+1})\Bigg\}+\frac{-\lambda}{4r^5}\,
 \Bigg\{ 2\,\dot F_{l}(l+1)^{2}(N-l-1)^{2} (1-\cos F_{l})^{2}-\dot F_{l-1}
\nonumber \\
&&l(l+1)(N-l)(N-l-1)(1-\cos F_{l-1})(1-\cos F_{l})-\dot F_{l+1}(l+1)(l+2)(N-l-1)\nonumber \\
&&(N-l-2)(1-\cos F_{l})(1-\cos F_{l+1})\Bigg\}
  +\frac{\lambda}{8r^4}\Bigg\{2\,\dot F_{l}^{2}(l+1)^{2}(N-l-1)^{2}
   (1-\cos F_{l})
\nonumber \\ 
&&\sin F_{l}-\dot F_{l-1}^{2}l(l+1)(N-l)(N-l-1)\sin F_{l-1}(1-\cos F_{l}) 
  -\dot F_{l+1}^{2}(l+1)(l+2)
\nonumber \\
&&(N-l-1)(N-l-2)(1-\cos F_{{l}})\sin F_{l+1}\Bigg\}=0.
\end{eqnarray}

When $N = 3$, the solution of the 2 equations lead to exact solutions of the 
model, while for larger values of $N$, the ansatz \Ref{ansatz} corresponds to 
low-energy configurations.

We would like to point out at this stage that as proved in \cite{sun}, 
the topological charge for the configuration \Ref{ansatz} is given by 
\begin{equation}
B = \sum_{i=0}^{N-2} {\cal D}_k (F_i-\sin F_i)_{r=0}^{r=\infty}
\end{equation}
where 
\begin{equation}
{\cal D}_k = -i \frac{1}{4 \pi^2} 
    \int \frac{\cmod{P^{k+1}_+ h}}{\cmod{P^k_+ h}} d\xi d\bar\xi 
\end{equation}
takes integer values given by the degree in $\xi$ of the wedge product 
\cite{Wojtek} of $h$ and its derivatives 
\begin{equation}
{\cal D}_k = \frac{1}{2\pi} deg(h^{(k)}) \qquad\qquad 
h^{(k)} = h \wedge \partial_{+} h \wedge ... \wedge 
\partial_{+}^k h\qquad k = 0, .. N-1.  
\end{equation}

Each configuration is thus characterised by the boundary conditions for the 
profile function $F_i$ and we can without loss of generality impose the 
condition 
$lim_{r \rightarrow \infty} F_i(r) = 0$. 
For the configuration to be well-defined at the origin we must also impose a
condition of the type
\begin{eqnarray}
F_i(0) &=& n_i\, 2 \pi 
\end{eqnarray}
where the $n_i \in N$. 

%%%%%%%%%%%%%%%%%%%%%%%%%%%%%%%%%%%%%%%%%%%%%%%%%%%%%%%%%%%%%%%%%%%%
\section{Radially symmetric $SU(3)$ Solutions}

To describe the solution of the $SU(3)$ model, we use the profile $F=F_0$ and
$g=F_1$ and the energy \Ref{SunEnergy} simplifies to

\begin{eqnarray}
\label{energyFg}
E&=&\fr{1}{6\pi}\int\, r^2 dr
\Bigg\{ 
   \frac{2}{3}\,(\dot{g}^2 + \dot{F}^2 + \dot{g}\,\dot{F})
+ \frac{1}{r^2} \left((1-\cos F) ((1-\lambda)\dot{F}^2+4) 
\right.\nonumber\\ 
&&\left.       +  (1- \cos g) ((1-\lambda)\dot{g}^2 +4)
                    \right) 
     + (1-\lambda) \frac{2}{r^4}
         \Big( (1-\cos F)^2   
                 - (1-\cos F) (1-\cos g) 
\nonumber\\
&&
+ (1-\cos g)^2 \Big) 
  + \frac{\lambda}{2 r^4}
        \left(\dot{F}^2\,(1-\cos F)^2
        + \dot{g}^2 (1-\cos g)^2 
         -  (1-\cos F) (1-\cos g) \dot{g}\dot{F}
       \right)
\Bigg\}.\nonumber\\
&&
\end{eqnarray}
The equations for the profile function $F$ and $g$ are then given by
\begin{eqnarray}
\label{eqn_g}
&& g_{rr} + \frac{1}{2} F_{rr}\,+ \frac{F_r}{r} + 2 \frac{g_r}{r}
+\frac{3}{2 r^2} \Big((1-\lambda)(1-\cos g) g_{rr} 
                      +\frac{1}{2} \sin g ((1-\lambda)g_r^2 -4) 
                 \Big)\nonumber\\
&&                      +\frac{1}{2} \sin g ((1-\lambda)g_r^2 -4) 
 +(1-\lambda) \frac{3}{2 r^4} 
           \big((1- \cos F) - 2 (1- \cos g)\big)\sin(g)
\nonumber\\
&&    +\frac{3 \lambda}{8 r^4} (1-\cos g)
    \Bigg(2 \big(\sin g  g_r^2 
        + (1-\cos g) (g_{rr}- 2 \frac{g_r}{r})\big)
\nonumber\\
&&        - \sin F  F_r^2 -(1-\cos F)(F_{rr}-2\frac{Fr}{r})
    \Bigg) = 0
\end{eqnarray}
\begin{eqnarray}
\label{eqn_F}
&&F_{rr}+ \frac{1}{2}\,g_{rr} + 2\,\frac{F_r}{r} + \frac{g_r}{r} 
 + \frac{3}{4 r^2}
    \left(\sin F ((1-\lambda)F_r^2-4) + 2 (1-\lambda)(1-\cos F ) F_{rr}\right)
\nonumber\\
&&   - (1-\lambda)\frac{3}{2 r^4}
          (2(1-\cos F)-(1-\cos g) )\sin F
  +\frac{3 \lambda}{8 r^4} (1-\cos F) \Bigg(12 (\sin F\, F_r^2
\nonumber\\
&&        +(1-\cos F)(F_{rr} - 2\frac{F_r}{r})) -\sin g\, g_r^2 
   - (1-\cos g) (g_{rr}-2 \frac{g_r}{r}))\Biggr) = 0.
\end{eqnarray}
The topological charge of the solution now reads 
\begin{eqnarray}
&&B= \fr{1}{\pi} \Bigg( 
              \left(F-\sin F\right) \vert_{r=0}^{r=\infty} 
               + (g -\sin(g))\vert_{r=0}^{r=\infty}\Bigg)
\end{eqnarray}
and if we take the boundary conditions 
\begin{eqnarray}
F(0) &=& n_F 2 \pi \nonumber\\ 
g(0) &=& n_g 2 \pi 
\end{eqnarray}
where $n_F$ and $n_g$ are integers, we have $B= 2 (n_f+n_g)$.
When $n_F$ and $n_g$ are of opposite signs, we can interpret the solutions 
as a mixture of Skyrmions and anti Skyrmions.

In Table 1, we give the energy of the hedgehog solution ($B=1$) for the 
$SU(2)$ model. This solution is an embedded solution of any $SU(N)$ model 
and it is the solution with the lowest energy. We thus use it as the 
reference energy for all the other solutions. 

In Table 2 we present the properties of the different solutions for the
$SU(3)$ models. The first two columns specify the boundary condition of 
the solution, and the third columns gives the topological charge of that 
solution. In column 4 and 5 we give the energy of the solutions for the pure
Skyrme model and the pure Sk6 model while column 6 and 7 give the 
corresponding relative energy per Skyrmion, that is the energy divided by
the energy of the single Skyrmion and the total number of Skyrmions. 
For the solutions corresponding to the superposition of Skyrmions and 
anti-Skyrmion, we define the total number of Skyrmions as the total number of 
Skyrmions and anti-Skyrmions. Notice that the cases $n_g=0, n_F=1$ and 
$n_g=1, n_F = 0$ correspond to the same solution modulo an internal rotation.

In Figure 3, we present the energy of the 3 different types of solution
as a function of $\lambda$.

\vskip 5mm
\begin{figure}[htbp]
\unitlength1cm \hfil
  \includegraphics[width=8cm]{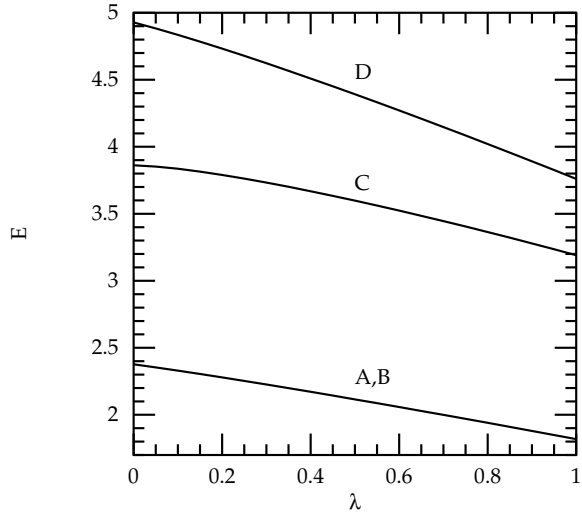}
\caption{Energy of the $SU(3)$ solution for the boundary conditions 
(A) $n_F=0, n_g=1$, (B) $n_F=1, n_g=0$,
(C) $n_F=1, n_g=-1$, (D) $n_F=1, n_g=1$.}
\end{figure}

\begin{table}[htbp]
\begin{center}
\begin{tabular}{|c|c||c|c| }
\hline
\multicolumn{2}{|c||}{$SU(2)$} & \multicolumn{2}{c|}{Energy} \\
\hline & & & \\ 
$n_g$ & $B$ & $E(0)$ & $E(1)$\\
\hline & & & \\
 1 & 1 & 1.2315 & 0.9395\\
\hline
\end{tabular}
\end{center}
\caption{Topological charge and Energy of the hedgehog $SU(2)$ solution.}
\end{table}

\begin{table}[htbp]
\begin{center}
\begin{tabular}{|c|c|c||c|c||c|c| }
\hline\multicolumn{3}{|c||}{$SU(3)$} & \multicolumn{2}{c||}{Total Energy}
& \multicolumn{2}{c|}{Relative Energy} \\
\hline & & & & & & \\ 
$n_F$ & $n_g$ & $B$ & $E(0)$ & $E(1)$&$E_B(0)/(|B| E_1(0))$ & $E_B(1)/(|B| E_1(1))$\\
\hline & & & & & & \\
1 & 1 & 4 & 4.928 & 3.758    & 1      & 1\\
1 & 0 & 2 & 2.377 & 1.819    & 0.965  &  0.968  \\
0 & 1 & 2 & 2.377 & 1.819    & 0.965  &  0.968 \\
1 & -1& 2-2 & 3.862& 3.191   & 0.784  &  0.849 \\
\hline
\end{tabular}
\end{center}
\caption{Topological charge and Energy of some $SU(3)$ solutions.}
\end{table}

%%%%%%%%%%%%%%%%%%%%%%%%%%%%%%%%%%%%%%%%%%%%%%%%%%%%%%%%%%%%%%%%%%%%%%%%%%%%
\section{Low Energy $SU(4)$ Configurations}

As was shown in the last two sections, the ansatz \Ref{ansatz} provides
an exact solution of the sixth order model only for the $SU(3)$ model, or
when $\lambda=0$, that is for the usual Skyrme model. For the $SU(N)$ model 
with $N \ge 4$, the ansatz still produces low-energy configurations. In 
particular, when $\lambda$ is small, we can expect the ansatz to be very close 
to an exact solution. In this section we look at some configurations of the
$SU(4)$ model. For this model, we have three profile functions $F_0$, $F_1$ and
$F_2$ and the energy for the general ansatz \Ref{ansatz} is explicitly 
given by

\begin{eqnarray}
E\!\!\!\!&=&\!\!\!\frac{1}{6\pi}\int\!\! r^{2} dr\Bigg\{
{1\over 4}\left(3\dot F_0^2+4\dot F_1^2+3\dot F_2^2+4\dot F_0\dot F_1+
                4\dot F_1\dot F_2+2\dot F_0\dot F_2\right)+
{2\over r^2}\left[3(1\!-\!\cos F_0)\!\right.
\nonumber\\
&&\left.
            +\!4(1\!-\!\cos F_1)\!+\!3(1\!-\!\cos F_2)\right]
+(1-\lambda)\Bigg\{{1\over 2 r^2}\left[3 \dot F_0^2(1\!-\!\cos F_0)\!+
      \!4 \dot F_1^2 (1\!-\!\cos F_1)\!+\right.
\nonumber\\
&&\left.
      \!3 \dot F_2^2 (1\!-\!\cos F_2)\right]
      +\fr{1}{2r^4}\,\{9\,(1-\cos F_0)^2+16\,(1-\cos F_1)^2+9\,(1-\cos F_2)^2
\nonumber\\
&&
- 12 (1-\cos F_0) (1-\cos F_1)-12(1-\cos F_1)(1-\cos F_2)\}\Bigg\}\nonumber\\
&&
+\fr{\lambda}{8 r^4}\,\{ 
  9 \dot F_0^2 (1-\cos F_0)^2 + 16 \dot F_1^2 (1-\cos F_1)^2
 +9 \dot F_2^2 (1-\cos F_2)^2 \nonumber\\
&&
 - 12 F_0 F_1 (1-\cos F_0)(1-\cos F_1) - 12 F_1 F_2 (1-\cos F_1)(1-\cos F_2)
\}
\Bigg\}
\label{SU4energy}
\end{eqnarray}
from which we can derive the following equations
\begin{eqnarray}
&&\left(\frac{3\lambda (1-\cos F_0)^2}{2r^4}
 +\frac {2(1-\lambda)(1-\cos F_0)}{r^2}+1\right){\ddot F}_{0}
 +\left(\frac{2}{3}-\frac {\lambda (1-\cos F_0)(1-\cos F_1)}{r^4}\right) 
         {\ddot F}_{1}
\nonumber \\
&&+\,\frac{1}{3}{\ddot F}_{2}\,-\,\frac {4\sin F_0}{r^2}\,
+\,\frac {6\,{\dot F}_0+4{\dot F}_1+2{\dot F}_2}{3 r}\,
+\,\frac {(1-\lambda){\dot F}_{0}^{2}\sin F_0}{r^2}
\nonumber \\
&&
   +(1-\lambda) \frac {\sin F_0}{r^4}(4(1-\cos F_1)-6(1-\cos F_0))
+\lambda\frac {(1-\cos F_0)}{r^4}
    \left(\frac{3}{2}\,{\dot F}_0 ^{2}\sin F_0
          -\,{\dot F}_{1}^{2}\sin F_1\right)
\nonumber \\
&&-\lambda\frac{(1-\cos F_0)}{r^5}
      \left(3{\dot F}_0(1-\cos F_0) -2\,{\dot F}_1(1-\cos F_1)\right)=0,
\end{eqnarray}

\begin{eqnarray}
&&\left (\frac{1}{2} -\frac {3\lambda(1-\cos F_0)(1-\cos(F_1)}{4 r^4}\right)
         {\ddot F}_0
  +\left(1+\frac {2\lambda (1-\cos F_1)^2}{r^4}
          +\frac {2(1-\lambda)(1-\cos F_1)}{r^2}\right) {\ddot F}_1 
\nonumber \\
&&+\left(\frac{1}{2} -\frac {3\lambda(1-\cos F_1)(1-\cos F_2)}{4 r^4}\right )
  {\ddot F}_2+
  {\frac {(1-\lambda){\dot F}_{1}^{2}\sin F_1}{r^2}
  +\frac {\dot F_0+2\,{\dot F}_1+{\dot F}_2}{r}-4\frac {\sin F_1}{r^2}}
\nonumber \\
&& +(1-\lambda) \frac {\sin F_1}{r^4}
     \left(3 (1-\cos F_0)+3 (1-\cos F_2)-8(1-\cos F_1)\right)
\nonumber \\
&& -\frac{\lambda}{r^5} (1-\cos F_1)
\left (4\,{\dot F}_1(1-\cos F_1)
     -\frac{3}{2}\,{\dot F}_0 (1-\cos F_0)
     -\frac{3}{2}\,{\dot F_2} (1-\cos F_2)\right )
\nonumber \\
&&+\frac {\lambda}{r^4}(1-\cos F_1) 
 \left (2\,{\dot F}_1^{2}\sin F_1
        -\frac{3}{4} {\dot F}_0^{^2}\sin F_0 
        -\frac{3}{4} {\dot F}_2^{2}\sin F_2\right ) = 0
\end{eqnarray}
and
\begin{eqnarray}
&&\left (\frac{2}{3}-\frac {\lambda(1-\cos F_1)(1-\cos F_2)}{r^4}\right)
 {\ddot F}_1
  +\left (\frac {3\lambda(1-\cos F_2)^{2}}{2r^4}
  +{\frac {2(1-\lambda)(1-\cos F_2)}{r^2}
  }+1\right ){\ddot F}_2 
\nonumber  \\
&&+\frac{1}{3}{\ddot F}_0
      + \frac {2{\dot F}_0+4{\dot F}_1+6\,{\dot F}_2}{3 r}
      -4\frac{\sin F_2}{r^2}
 +\frac {(1-\lambda){\dot F}_2^{2}\sin F_2}{r^2}
+(1-\lambda)\frac {\sin F_2}{r^4}\left(4(1-\cos F_1)\right.\nonumber \\
&&\left.
-6(1-\cos F_2)\right)
 -\lambda \frac{(1-\cos F_2)}{r^5}
   \left (3\,{\dot F}_2(1-\cos F_2) -2\,{\dot F}_1 (1-\cos F_1) \right)
\nonumber \\
&&
+\lambda\frac{(1-\cos F_2)}{r^4} 
   \left(\frac{3}{2}\,{\dot F}_2^{2} \sin F_2 
         -\,{\dot F}_1^{2}\sin F_1\right )=0.
\end{eqnarray}
Describing the boundary condition for the profile functions as before, 
$F_i(0) = n_i 2 \pi$, the topological charge is given by
\begin{equation}
B = 3 n_0 + 4 n_1 + 3 n_2.
\end{equation}

In Table 3 we present the energy values of various types of configurations
when $\lambda=0$ and $\lambda=1$. We notice that when $\lambda=0$, 
the solutions are symmetric under the exchange $f_0 \leftrightarrow f_2$, but 
that the sixth order term breaks the symmetry. This results in a difference of
energy between the configuration with $n_0=0, n_1=0, n2=1$ and 
$n_0=1, n_1=0, n2=0$ as well as between the configurations with 
$n_0=1, n_1=1, n2=0$ and $n_0=0, n_1=1, n2=1$. 
In Figure 4, we present the curve for the energy of the
configurations as a function of $\lambda$.

\begin{table}[htbp]
\begin{center}
\begin{tabular}{|c|c|c|c||c|c||c|c| }
\hline\multicolumn{4}{|c||}{$SU(4)$} & \multicolumn{2}{c||}{Total Energy}
& \multicolumn{2}{c|}{Relative Energy} \\
\hline & & & & & & & \\ 
$n_0$ & $n_1$ &$ n_2$ & $B$ & $E(0)$ & $E(1)$&$E_B(0)/(|B| E_1(0))$ & $E_B(1)/(|B| E_1(1))$\\
\hline & & & & & & &\\
0 & 0 & 1 & 3 & 3.51739 & 2.66653 & 0.95210 & 0.94598 \\
1 & 0 & 0 & 3 & 3.51739 & 2.72915 & 0.95210 & 0.96819 \\
0 & 1 & 0 & 4 & 4.78807 & 6.33322 & 0.97204 & 1.68507 \\
1 & 0 & 1 & 6 & 7.22464 & 6.04604 & 0.97780 & 1.07244 \\
1 & 1 & 0 & 7 & 8.45219 & 6.62998 & 0.98052 & 1.00802 \\
0 & 1 & 1 & 7 & 8.45219 & 7.28058 & 0.98052 & 1.10694 \\
1 & 1 & 1 & 10 & 12.311 & 9.39605 & 1 & 1\\
\hline
\end{tabular}
\end{center}
\caption{Topological charge and Energy of some $SU(4)$ configurations.}
\end{table}

\vskip 5mm
\begin{figure}[htbp]
\unitlength1cm \hfil
  \includegraphics[width=8cm]{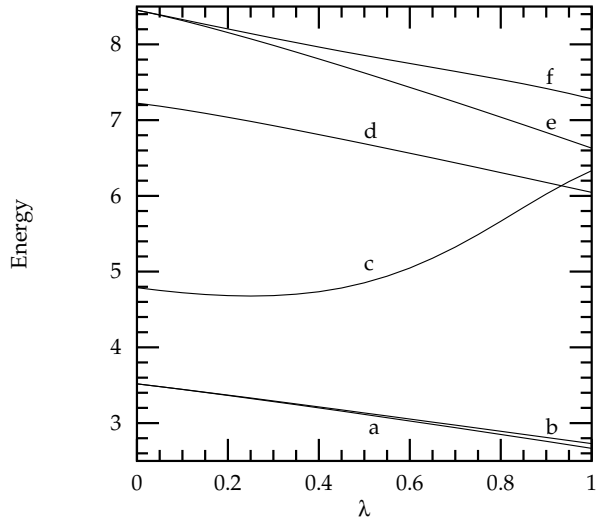}
\caption{Energy density of the $SU(4)$ multi-projector ansatz 
(a) $n_0=0, n_1=0, n_2=1$; (b) $n_0=1, n_1=0, n_2=0$; 
(c) $n_0=0, n_1=1, n_2=0$; (d) $n_0=1, n_1=0, n_2=1$; 
(e) $n_0=1, n_1=1, n_2=0$; (f) $n_0=0, n_1=1, n_2=1$.
} 
\end{figure}

%%%%%%%%%%%%%%%%%%%%%%%%%%%%%%%%%%%%%%%%%%%%%%%%%%%%%%%%%%%%%%%%%%%%%%%%%%
\section{$SU(N)$ Low Energy configuration }

After inserting the ansatz \Ref{genansatz} in the full equation for the
$SU(N)$ model, we found that we had only two independent profile functions 
$g_0$ and $g_1$ and that the ansatz would only provide solutions for the 
$SU(3)$ model. One can nevertheless use the $SU(N)$ ansatz to compute low 
energy configurations. For example if we consider the reduced ansatz 
defined by \Ref{genansatz} together with the constraint $g_i = g_{i+2}$
and define the profiles $F=g_0-g_1$ and $g=g_{N-2}$ we can minimise
the energy \Ref{SunEnergy} and solve the equations for $F$ and $g$ for
various boundary conditions. We found that to get configurations corresponding 
to a bound state, {\it i.e.} a configuration with an energy per Skyrmion 
smaller than the energy of the hedgehog solution, we must take $n_F=0$ and 
$n_g=1$. The energies that we found are given in Table 3. 

\begin{table}[htbp]
\begin{center}
\begin{tabular}{|c|c||c|c||c|c| }
\hline\multicolumn{2}{|c||}{} & \multicolumn{2}{c||}{Total Energy}
& \multicolumn{2}{c|}{Relative Energy} \\
\hline & & & & & \\ 
Model& $B$ & $E(0)$ & $E(1)$&$E_B(0)/(|B| E_1(0))$ & $E_B(1)/(|B| E_1(1))$\\
\hline & & & & & \\
$SU(3)$ & 2 & 2.377 & 1.819    & 0.965  &  0.968 \\
$SU(4)$ & 3 & 3.624 & 2.759  & 0.981 & 0.979 \\
$SU(5)$ & 4 & 4.811 & 3.632  & 0.977 & 0.966 \\
$SU(6)$ & 5 & 6.015 & 4.518 & 0.977 & 0.962 \\
\hline
\end{tabular}
\end{center}
\caption{Topological charge and energy for the reduced ansatz with $n_F=0$ and 
$n_g=1$.}
\end{table}

\vskip 5mm
\begin{figure}[htbp]
\unitlength1cm \hfil
  \includegraphics[width=8cm]{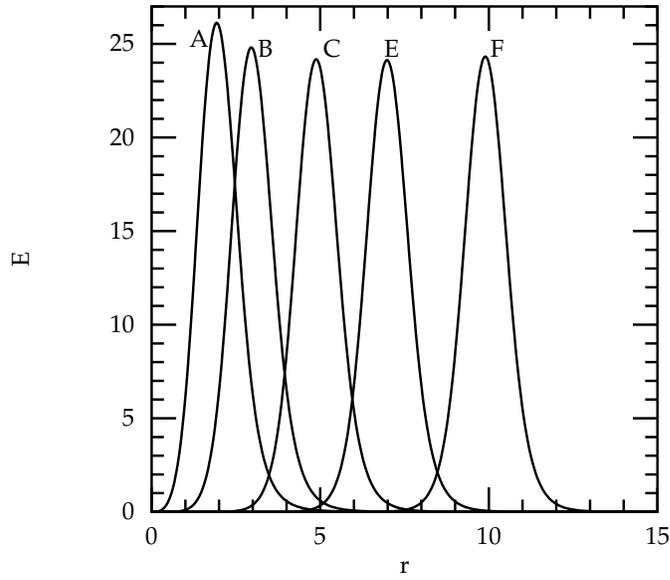}
\caption{Energy density of the multi-projector 
solution with $n_F=0$, $n_g=1$, $\lambda=0.5$. 
(A) N=10, (B) N=20, (C) N=50, (D) N=100,  (E) N=200.} 
\end{figure}

\vskip 5mm
\begin{figure}[htbp]
\unitlength1cm \hfil
 \centering
 \subfigure[]{
   \includegraphics[width=7cm]{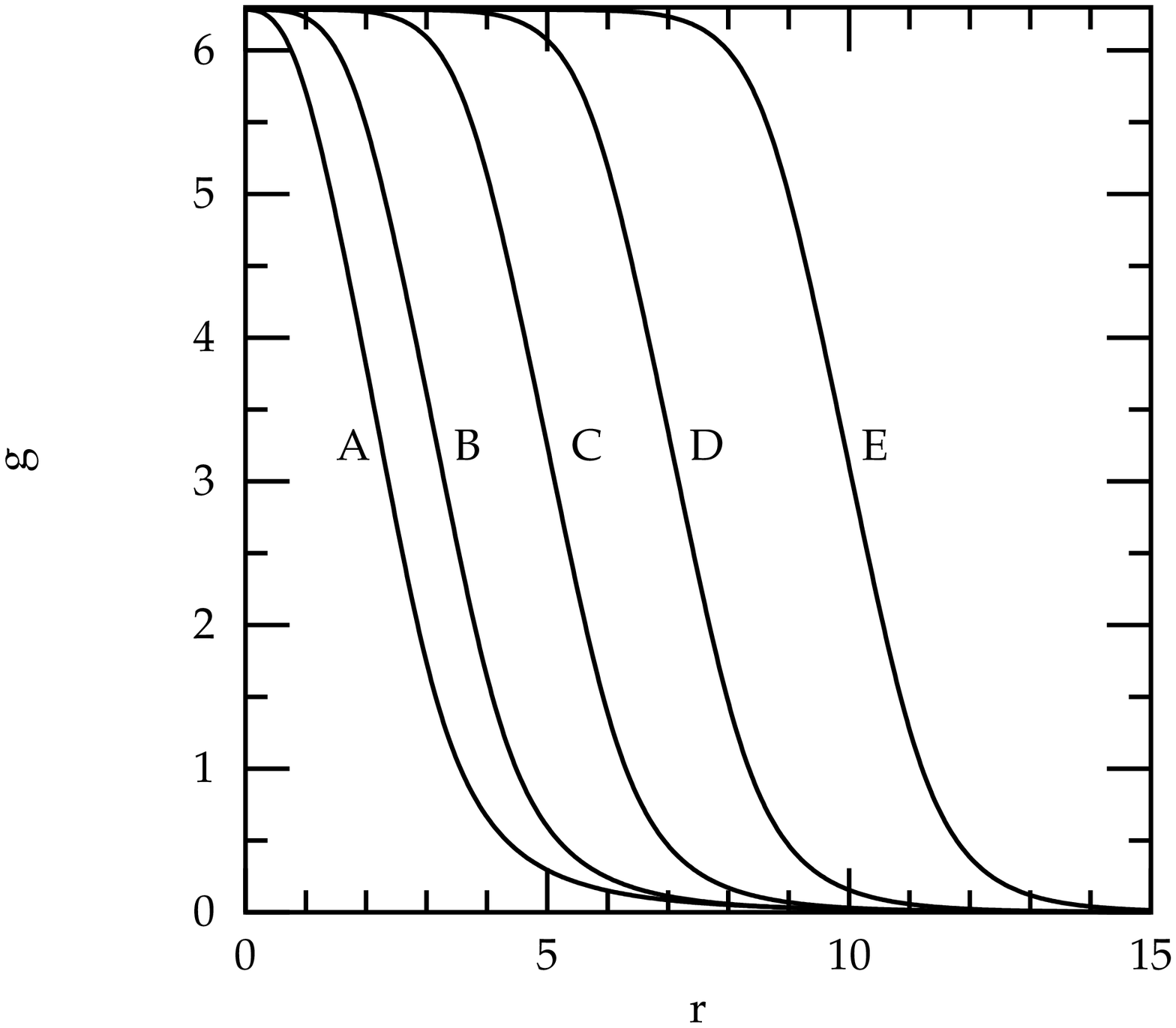}
 }
 \qquad
 \subfigure[]{
 \includegraphics[width=7cm]{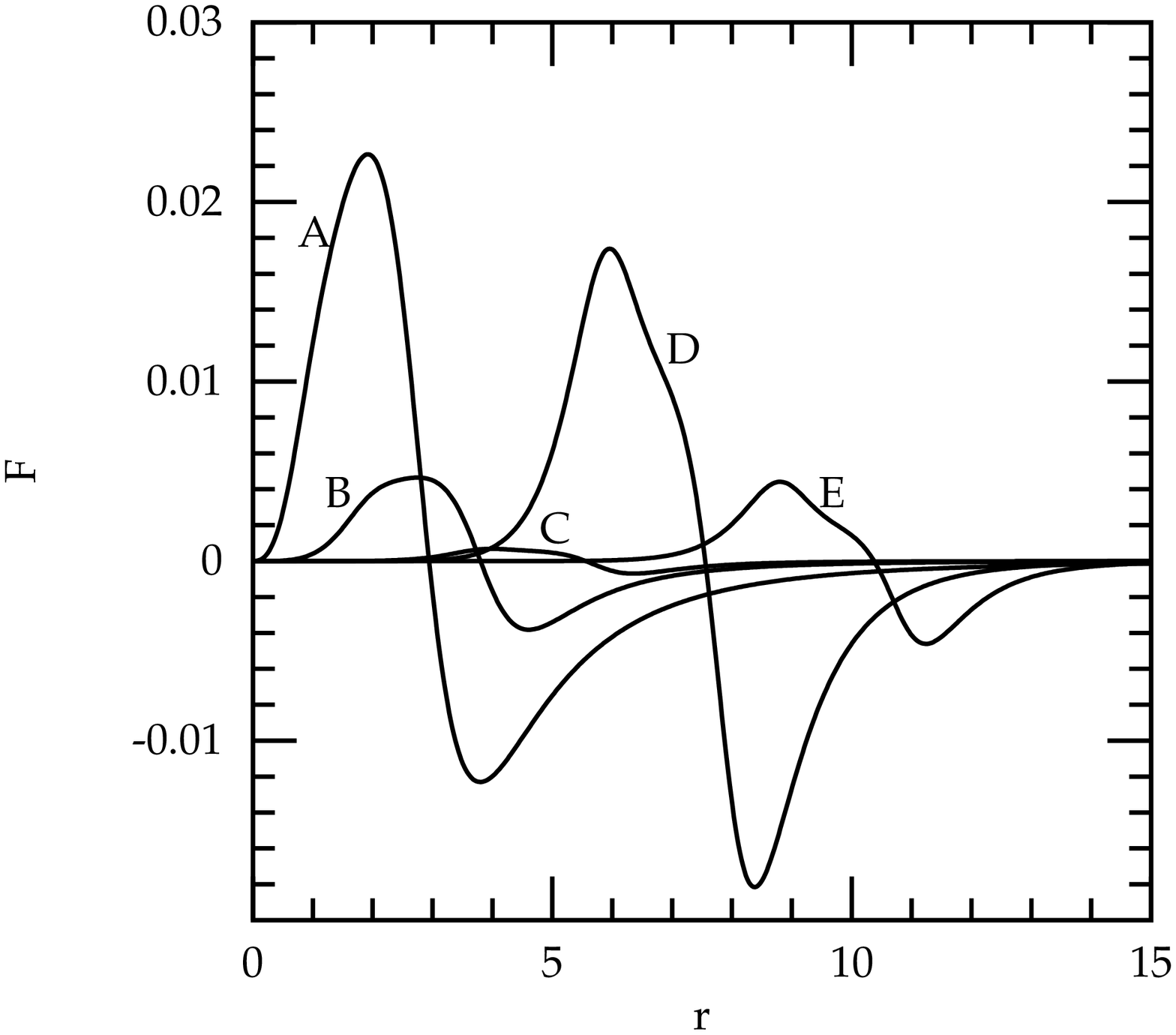}
 } 
\caption{Profile (a) $g$ and (b) $F$ of the multi-projector solution with 
$n_F=0$, $n_g=1$, $\lambda=0.5$. 
(A) $F$ for N=10, (B) $F$ for N=20, (C) $F$ for N=50, 
(D) $100\times F$ for N=100,  (E) $100\times F$ for N=200.} 
\end{figure}

In Figures 5 and 6 we present the profile and the energy density for different 
values of $N$ and for $\lambda=0.5$. It shows that the energy density
has the shape of a hollow sphere of radius $r=0.7 \sqrt{N}$. The profile
$g$ has the same shape for all values of $N$ but is shifted to the right 
as $N$ increases. The profile $F$ on the other hand is also shifted as the
shell radius increases, but its amplitude decreases like $1/N^2$; note that 
in Figure 6, the profile for $N=100$ and $N=200$ have been multiplied by $100$
to make them visible.
For other values of $\lambda$ the graphics look very much the same except
that the shell radius and width are slightly different, but the conclusions 
remain the same.

Figure 6.b suggests to simplify the ansatz further for large $N$
by taking $F(r)=0$. This implies that
$g_i=g$ $\forall i$ and the multi-projector ansatz \Ref{genansatz} 
becomes
\begin{equation}
\label{limitansatz}
U = \mbox{exp}\left(- i g (P_{N-1}- {\it I}/N)\right)
\end{equation}
where $P_{N-1}$ can also be written as 
\begin{equation}
P_{N-1} = \frac{\tilde h \tilde h^\dagger}{\cmod{\tilde h}}
\end{equation}
where $\tilde h$ is equal, up to a unitary rotation, to the 
complex conjugate of the holomorphic vector $V_0$ defined in 
(\ref{V}-\ref{fk}) : $\tilde h = A \bar V_0$ for some $A \in SU(N)$ 
with $\partial_\xi A = \partial_{\bar\xi}A = 0$.
This is shown by using the fact that $P_{N-1}$ is an anti-holomorphic 
projector \cite{Wojtek} and that solving \Ref{normV} recursively we have
\begin{equation}
\cmod{V_{k}} = \frac{k! (N-1)!}{(N-1-k)!} \mod{1+\cmod{\xi}}^{N-1-2k} 
\end{equation}
and so $\cmod{V_{N-1}} = (N-1)!^2\mod{1+\cmod{\xi}}^{1-N}$. 
Knowing that up to an 
overall coefficient $\cmod{V_{N-1}}$ is a polynomial in $\bar\xi$ of degree
$N-1$, we can conclude that up to a unitary iso-rotation, $V_{N-1}$ is equal 
to the complex conjugate of $V_0$.

The topological charge of the anti-holomorphic projector $P_{N-1}$ is
equal to $1-N$ and as the profile function is $-g$, the baryon number for
this configuration is $N-1$. The ansatz \Ref{limitansatz} is not a solution, 
but its energy 
\begin{eqnarray}
\label{energyLA}
E&=&\fr{1}{6\pi}\int\, r^2 dr
\Bigg\{ 
   \frac{N-1}{N}\,\dot{g}^2 + \frac{1}{2 r^2}  
     +  (N-1)(1- \cos g) ((1-\lambda)\dot{g}^2 +4)
                    \nonumber\\
&& + \frac{1}{2 r^4} (N-1)^2 (1-\cos g)^2 
     \left((1-\lambda) + \frac{\lambda}{4 r^4}\dot{g}^2 \right)  
\Bigg\}
\end{eqnarray}
can easily be computed by solving the equation
\begin{eqnarray}
 \label{eqn_gLA}
&&2 g_{rr} + 4 \frac{g_r}{r} +\frac{N}{r^2} 
      \Big((1-\lambda)(1-\cos g) g_{rr}
            +\frac{1}{2} \sin g ((1-\lambda)g_r^2 -4) 
      \Big)
\nonumber\\
&&  +\frac{\lambda}{4 r^4} N (N-1)(1-\cos g)
    (\sin g  g_r^2 + (1-\cos g) (g_{rr}- 2 \frac{g_r}{r}))
    = 0.
\end{eqnarray}

\vskip 5mm
\begin{figure}[htbp]
\unitlength1cm \hfil
\begin{center}
 \includegraphics[width=8cm]{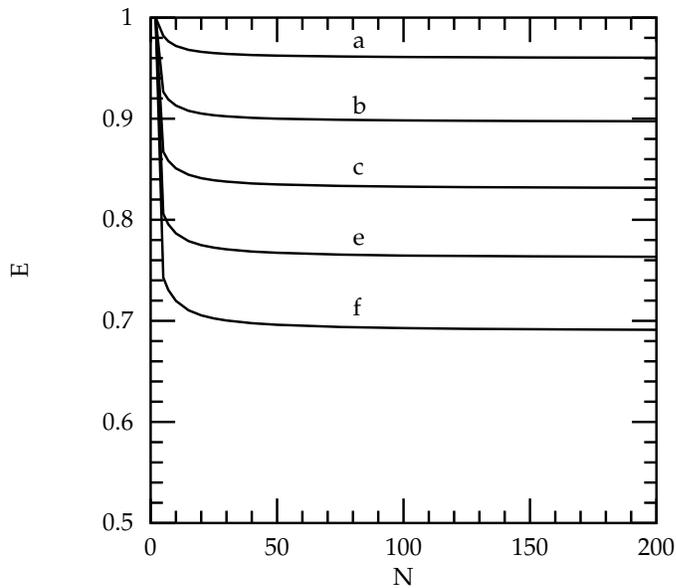}
\end{center}
\caption{Energy $E/(E_{B=1} (N-1))$ of the $SU(N)$, configuration 
\Ref{limitansatz}
for (a) $\lambda=0$, (b) $\lambda=0.25$,  (c) $\lambda=0.5$,  
(d) $\lambda=0.75$, (e) $\lambda=1$}
\end{figure}

In Figure 7, we present the relative energy,
$E(\lambda)/(E_{B=1}(\lambda) (N-1))$, of this configuration as a 
function of $N$ for different values of $\lambda$. We see that this 
configuration corresponds to a bound state of Skyrmions and that the energy 
per Skyrmion decreases with $N$. The energy of this configuration
corresponds to an upper bound for the energy of the $B=N-1$ radially symmetric 
solution of the $SU(N)$ model and these configurations
correspond to bound states of Skyrmions for all values of $N$ and
all values of $\lambda$. As every $SU(p)$ solution can be trivially embedded
in an $SU(q)$ solution when $p \le q$ we can claim that for every $B < N$ 
the $SU(N)$ model has a radially symmetric solution of charge $B$ corresponding 
to a bound state. With the exception of the hedgehog solutions, these solutions
are expected to be unstable when the radial symmetry is broken as their 
energies are larger than the known $SU(2)$ solutions \cite{FP1}.

\section{Conclusions}

In this paper we have shown how to construct some radially symmetric solutions
of the $SU(3)$ sixth order Skyrme model. The construction is similar to the 
one used for the pure Skyrme model in \cite{sun} except that, because of an 
extra constraint, the construction only works for the $SU(3)$ model. 
The same ansatz can nevertheless be used to compute low-energy configurations
of the $SU(N)$ model. In particular we showed that for every $N$ there is a 
radially symmetric solution of charge $B < N$ which corresponds to a bound 
state of Skyrmion. 
 
\section{Acknowledgement}
We would like to thank W.J. Zakrzewski for useful discussions during this 
work.


\begin{thebibliography}{99}
\bibliographystyle{plain}


\bibitem{sun}
T.Ioannidou, B.Piette and W.J.Zakrzewski, J.Math.Phys.
{\bf 40:6353-6365} (1999)

\bibitem{Skyr1}
T. H. R. Skyrme, Nucl. Phys. {\bf 31}, 556 (1962).

\bibitem{Hooft}
G.'t Hooft, Nucl. Phys. {\bf B72}, 461 (1974)

\bibitem{Witt1}
E.Witten, Nucl. Phys. {\bf B160}, 57 (1979)

\bibitem{Witt2}
E.Witten, Nucl. Phys. {\bf B223} , 433 (1983)

\bibitem{Jackson}
A.Jackson, A.D.Jakson, A.S.Goldhaber, \\ G.S.Brown and
L.C.Castillo, Phys. Lett. {\bf B154} , 101 (1985)

\bibitem{Marl1}
L.Marleau, Phys. Lett. {\bf B244}, 580 {1990}

\bibitem{Marl4}
L. Marleau, Phys. Rev.  {\bf D 45}, 1776 (1992)

\bibitem{Marl5}
L. Marleau, Phys. Rev.  {\bf D 63}, 036007 (2001)

\bibitem{HA}
M.A. Halasz and R.D. Amado, Phys. Rev. {\bf D 61}, 074022 (2000), 
Phys. Rev. {\bf D 63}, 054020 (2001)

\bibitem{Manton}
C. J. Houghton, N. S. Manton and P. M. Sutcliffe, Nucl. Phys. B
{\bf 510}, 507 (1998).

\bibitem{spherical}
T.Ioannidou, B.Piette and W.J.Zakrzewski, J.Math.Phys.
{\bf 40:6223-6233} (1999)

\bibitem{Hans}
H. J. Wospakrik, W. J. Zakrzewski J. Math. Phys. {\bf 42}, 1066 (2001)

\bibitem{FP1}
I. Floratos and B. Piette, hep-th/0103126 to appear in Phys. Rev. D.

\bibitem{Din}
A. Din and W.J. Zakrzewski, Nucl. Phys. {\bf B 174}, 397 (1980).

\bibitem{Wojtek}
W.J. Zakrzewski;{\it Low Dimensional Sigma Models} (IOP, Bristol 1989).



\end{thebibliography}
\end{document}